%% file: main.tex
\documentclass[sigconf]{acmart}

\input{preamble.tex}

\RemoveAlgoNumber
\AtBeginDocument{
  \providecommand\BibTeX{{
    \normalfont B\kern-0.5em{\scshape i\kern-0.25em b}\kern-0.8em\TeX}}}

\usepackage{mathtools}
\usepackage{graphicx}
\usepackage{balance} 
\usepackage[switch]{lineno}
\usepackage[ruled,vlined]{algorithm2e}
\usepackage{tabularx}
\usepackage{makecell}
\usepackage{subcaption}
\usepackage[utf8]{inputenc}
\usepackage[T1]{fontenc}
\usepackage{hyperref}
\usepackage{url}
\usepackage{booktabs}
\usepackage{amsfonts}
\usepackage{nicefrac}
\usepackage{microtype}
\usepackage{amsmath}

\setcopyright{acmcopyright}

\copyrightyear{2021} 
\acmYear{2021} 
\setcopyright{acmcopyright}\acmConference[ICAIF'21]{2nd ACM International Conference on AI in Finance}{November 3--5, 2021}{Virtual Event, USA}
\acmBooktitle{2nd ACM International Conference on AI in Finance (ICAIF'21), November 3--5, 2021, Virtual Event, USA}
\acmPrice{15.00}
\acmDOI{10.1145/3490354.3494386}
\acmISBN{978-1-4503-9148-1/21/11}

\begin{document}

\title[Learning to Classify and Imitate Trading Agents in Continuous Double Auction Markets]{Learning to Classify and Imitate Trading Agents in Continuous Double Auction Markets}

\author{Mahmoud Mahfouz}
\affiliation{
  \institution{J.P. Morgan AI Research, \\ Imperial College London}
  \city{London}
  \country{United Kingdom}
}
\email{mahmoud.a.mahfouz@jpmchase.com}

\author{Tucker Balch}
\affiliation{
  \institution{J.P. Morgan AI Research}
  \city{New York}
  \country{United States of America}
}
\email{tucker.balch@jpmchase.com}

\author{Manuela Veloso}
\affiliation{
  \institution{J.P. Morgan AI Research}
  \city{New York}
  \country{United States of America}
}
\email{manuela.veloso@jpmchase.com}

\author{Danilo Mandic}
\affiliation{
  \institution{Imperial College London}
  \city{London}
  \country{United Kingdom}
}
\email{d.mandic@imperial.ac.uk}

\renewcommand{\shortauthors}{Mahfouz et al.}

\input{body/_0_abstract}



\maketitle

\input{body/_1_introduction}
\input{body/_2_background}
\input{body/_3_contributions}
\input{body/_4_abm_lob}
\input{body/_5_classifying_trading_agents}
\input{body/_6_behavioural_cloning}
\input{body/_7_conclusion_future_work}

\newpage
\begin{acks}
    The authors would like to acknowledge our colleagues Thomas Spooner, Nelson Vadori, Leo Ardon, Pranay Pasula and Rui Silva for their input and suggestions at various key stages of the research.
    
    This paper was prepared for informational purposes in part by the Artificial Intelligence Research group of JPMorgan Chase \& Co and its affiliates ("J.P. Morgan"), and is not a product of the Research Department of J.P. Morgan. J.P. Morgan makes no representation and warranty whatsoever and disclaims all liability, for the completeness, accuracy or reliability of the information contained herein. This document is not intended as investment research or investment advice, or a recommendation, offer or solicitation for the purchase or sale of any security, financial instrument, financial product or service, or to be used in any way for evaluating the merits of participating in any transaction, and shall not constitute a solicitation under any jurisdiction or to any person, if such solicitation under such jurisdiction or to such person would be unlawful.
\end{acks}

\bibliographystyle{ACM-Reference-Format}
\bibliography{references}

\end{document}

%% file: body/_0_abstract.tex
\begin{abstract}
    Continuous double auctions such as the limit order book employed by exchanges are widely used in practice to match buyers and sellers of a variety of financial instruments. In this work, we develop an agent-based model for trading in a limit order book and show (1) how opponent modelling techniques can be applied to classify trading agent archetypes and (2) how behavioural cloning can be used to imitate these agents in a simulated setting. We experimentally compare a number of techniques for both tasks and evaluate their applicability and use in real-world scenarios.
\end{abstract}

%% file: body/_1_introduction.tex
\section{Introduction} \label{sec:introduction}

    The continuous double auction (CDA) is a well studied and heavily utilized mechanism for matching buyers and sellers in a variety of markets. It is typically used by trading exchanges in financial markets such as stocks and commodities exchanges. A CDA is referred to as a \emph{double} auction as it allows both buyers and sellers to submit orders and \emph{continuous} as trades are instantaneously executed upon receipt of compatible orders \cite{wellman2011trading}.
    
    The financial markets can be thought of as a multi-agent system composed of a variety of heterogeneous trading agents with different objectives and timescales. These agents also have different utility functions (\emph{e.g.} maximizing profit or reducing risk exposure) which are highly dependent on the behaviour and actions of the other market participants. Exchange-traded markets use a CDA mechanism and record the agents orders in a \emph{Limit Order Book}, which we outline in section \ref{subsec:lob}.
    
    In this paper, we explore the feasibility of classifying and imitating the behaviour of a representative set of trading agents in a limit order book. To this end, we attempt to answer the following questions:
    
    \begin{enumerate}
        \item Can we build a realistic agent-based model of a set of agent archetypes trading in a limit order book?
        \item Can we learn to classify these trading agent archetypes based on the limit order book data and agent orders?
        \item Can we imitate the behaviour of these trading agents using the limit order book data as inputs to a model and agent order information as the labels?
    \end{enumerate}
    
    Learning to classify and imitate trading agents is particularly useful for regulators around the world who are interested in preventing the occurrence of market manipulation. Moreover, it allows regulators to better understand the impact of various types of participants on the market which in turn leads to better regulations and stability in the financial markets.

%% file: body/_2_background.tex
\section{Background} \label{sec:background}

    \subsection{Limit Order Books} \label{subsec:lob}
       
        \begin{figure}[!htb]
             \centering
             \includegraphics[width=0.4\textwidth]{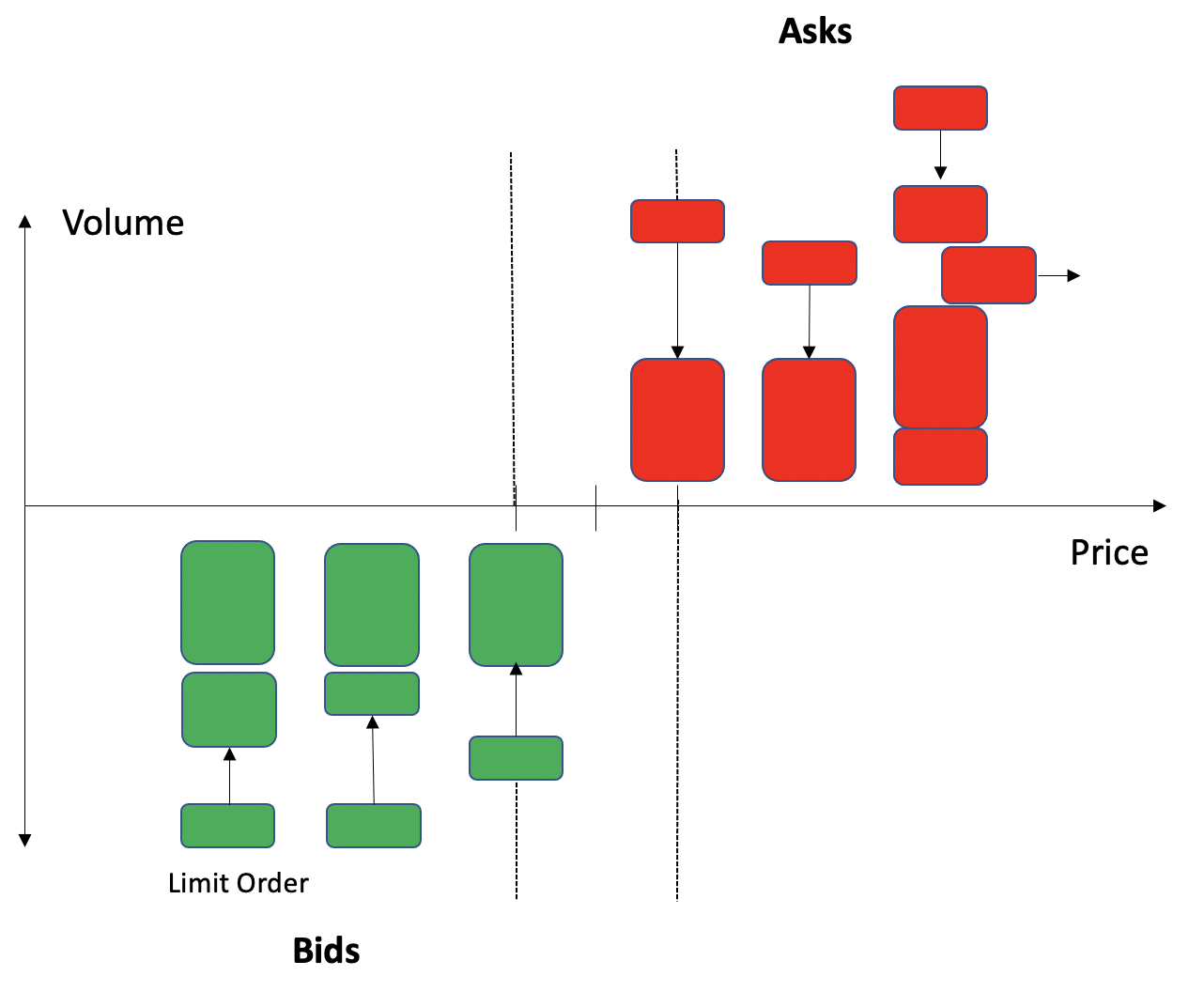}
             \caption{Visualization of the limit order book}
             \label{fig:LOB}
        \end{figure}
       
        The limit order book (LOB) is an indicator of the supply and demand for an exchange traded instrument at a given point in time (see figure \ref{fig:LOB}). It is an electronic record of all the outstanding buy and sell limit orders organized by price levels. The LOB is split into two sides; the ask and bid sides containing all the sell and buy limit orders respectively.
        
        A matching engine is used to match the incoming buy and sell orders. This typically follows the price/time priority rule \cite{preis2011price}, whereby orders are first ranked according to their price. Multiple orders having the same price are then ranked according to the time they were entered. If the price and time are the same for the incoming orders, then the larger order gets executed first. The matching engine uses the LOB to store pending orders that could not be executed upon arrival.
        
        Order types are distinguished between \emph{limit} orders and \emph{market} orders. A limit order includes a price that should not be exceeded in the case of a buy, or should not be gone below in the case of a sell. A market order indicates that the trader is willing to accept the best price available immediately. 
        
    \subsection{Opponent Modeling}
    
        Opponent modeling refers to a set of methods to construct models of the other agents in the same environment with the goal of predicting various properties of interest such as the agents' actions, types, beliefs and goals \cite{albrecht2018autonomous}. Different approaches are studied in the literature including policy reconstruction \cite{brown1951iterative, veloso1994planning}, type-based reasoning \cite{barrett2013teamwork, albrecht2015hba}, classification \cite{weber2009data}, plan recognition \cite{carberry2001techniques}, recursive reasoning \cite{carmel1996incorporating}, graphical models \cite{howard2005influence} and group modeling \cite{stone1999task}.
        
        Learning from data~\citep{shoham2008multiagent} is an attractive framework under which assumptions can be relaxed and flexible models can be developed and calibrated to real-world data. In particular, recent advances in machine learning~\citep{goodfellow2014generative, mnih2015human} allow for black-box, task-agnostic, large-scale modeling. However, in all these settings, the presence of multiple learning agents renders the training problem non-stationary and often leads to unstable dynamics or undesirable final results \cite{balduzzi2018mechanics, foerster2018learning}. Opponent modeling~\citep{carmel1995opponent} addresses this problem, attempting to \emph{decouple the sources of non-stationarity} by modeling separately the other learning agents from the stationary components of the environment.
    
    \subsection{Behavioural Cloning}
    
        Behavioural cloning or learning from demonstrations is a supervised imitation learning paradigm building a mapping from a historical set of observations to actions without recovering the reward function \cite{osa2018algorithmic}. The paradigm is particularly useful in applications whereby an agent is able to demonstrate its behaviour rather than specify its policy directly or formulate its reward function. 
        Behavioural cloning has been studied and applied in various domains including self-driving cars \cite{pomerleau1989alvinn}, playing the game Go \cite{silver2016mastering}, generating natural language \cite{wen2015semantically} and image captions \cite{karpathy2015deep}. \cite{osa2018algorithmic} provides an excellent overview of the key methods underpinning learning from demonstrations and current open research problems.
        
    \subsection{Related Work}

         The importance and applications of opponent modelling techniques in first-price sealed bid iterated auctions and continuous double auction markets is discussed in \cite{mahfouz2019importance}. \cite{yang2012behavior, yang2015gaussian} perform opponent modeling implicitly in a continuous double auction setup by using Inverse Reinforcement Learning (IRL) to characterize the different trading strategies in a limit order book. In their approach, they model the trades placed in the order book as a Markov Decision Process (MDP) and cluster the reward space to differentiate between high frequency, opportunistic and market making strategies.

%% file: body/_3_contributions.tex
\section{Our Contributions} \label{sec:contributions}
    
    \begin{enumerate}
        \item We build an agent-based model (ABM) consisting of four different trading agent archetypes following strategies common in practice and vary their behaviours to ensure a wide coverage of trading strategies. 
        \item We assess the realism of our market configuration and constituent agents by comparing the statistical properties of the simulated market to those observed in real markets.
        \item We formulate our approach to opponent modeling in a limit order book as a supervised learning problem of identifying (classifying) agent archetypes explicitly designed in the ABM. To this end, we designed a feed-forward neural network classifier and compared its performance against other various classification models.
        \item We show how a similar model to that developed for the opponent modelling task can be used for behavioural cloning and compare the distribution of actions between the predicted and ground truth labels. 
    \end{enumerate}

%% file: body/_4_abm_lob.tex
\section{Agent-based Model for trading in a Limit Order Book} \label{sec:abm_lob}

    Agent-based modeling (ABM) is a bottom-up approach to building synthetic multi-agent financial markets composed of various types of trading participants. In these models, the agents are reactive to new unseen scenarios allowing them to perform counterfactual reasoning and answer various what-if scenarios. In our work, we utilize ABM as a method to construct a synthetic limit order book market composed of various trading agent archetypes, which we attempt to classify and clone their respective behaviour.
    
    Game theory~\citep{shoham2008multiagent, wooldridge2009introduction} provides a mathematical formalism for describing the complex behaviours that emerge from the different agent interactions and suggests tools to develop successful strategies. Nonetheless, strong assumptions are traditionally made about having access to reliable models describing the system dynamics, which is invalid even in the simplest of setups. For example, in financial markets designing a strategy that takes into account the impact of its decisions, is as difficult as solving the original problem itself, and heuristic approaches~\citep{gsell2008assessing} are not adaptive hence their quality degrades as the dynamics drift over time. 
    
    \subsection{Simulator}
    
        We utilize and extend an open source discrete event simulation environment for our work \cite{byrd2019abides}. The simulation environment is designed for simulating trading agents in a continuous double auction market through a NASDAQ-like exchange agent with price-then-FIFO (first in, first out) matching rules. It also allows for the simulation of variable electronic network latency and agent computation delays, features that are specifically important for high frequency trading in exchange-traded markets. The agents communicate solely through a standardized message protocol similar to that used by NASDAQ. The environment is controlled by a simulation kernel which manages the flow of time and handles all the communication between the different agents in the simulation to request latest limit order book information, place market or limit orders, and cancel outstanding limit orders, etc.

    \subsection{Agents}
        
        Our agent population consists of a mixture of zero-intelligence (ZI) and strategic limited-intelligence agent archetypes. The ZI agents were originally introduced by \cite{gode1993allocative} to describe a family of trading agents submitting orders in the market in a random fashion without basing the decisions on knowledge of the current state of the limit order book. Interestingly, it was shown in this work that the allocative efficiency of a market arises from its structure and not the particular strategy or intelligence of its participants. 
        
        The strategic limited-intelligence agents in our work are based on the authors knowledge of the types of trading agents acting in real exchange-traded markets. We refer to the zero-intelligence agents as background agents and define three archetypes of strategic limited-intelligence agents: market makers, market takers and directional traders which we explain in detail in this section.

        \subsubsection{\textbf{Background Agents:}}
        
            Our background ZI agents consist of two agent types: (1) \emph{Noise} agents which closely resemble the agents introduced in \cite{gode1993allocative} and (2) \emph{Value} agents which base their decisions on an exogenous price time series.

            The noise agents represent traders participating in the market once a day \cite{kyle1985continuous} by placing a market order and their participation time is sampled from a $U$-quadratic distribution over the interval $[t_{\text{open}}, t_{\text{close}}]$, where $t_{\text{open}}$ and $t_{\text{close}}$ are the start and end of the trading day, respectively. This is designed to achieve higher trading activity at the beginning and at the end of a given trading day reflecting traders' propensity to be more active at these times, a pattern observed in real limit order books \cite{Dacorogna1993AGM}.

            The value agents trade based on an exogenous time series representing the agents' belief of the true fundamental value of the asset they are trading \cite{wang2017spoofing, wah2017welfare}. In our work, we define this time series to be the historical intra-day price of a specific stock trading in NASDAQ.
            
            \begin{algorithm}
                \DontPrintSemicolon
                \caption{Noise Agent trading strategy}
                \label{alg:noise_agent}
                \KwData{quantity $q$, simulation start time $t_{\text{sim\_start\_time}}$, simulation end time $t_{\text{sim\_end\_time}}$, market open time $t_{open}$, market close time $t_{close}$}
                \SetKwProg{noiseagent}{NoiseAgent}{}{end}
                \noiseagent{$(q, t_{\text{sim\_start\_time}}, t_{\text{sim\_end\_time}}, t_{open}, t_{close})$}{
                Sample $t \sim \quaddist(t_{open}, t_{close})$\;
                \eIf{$t \in [t_{\text{sim\_start\_time}}, t_{\text{sim\_end\_time}}]$\ }{\wakeup$(t)$\;}{\Return}
                Sample $\epsilon \sim \udist \{+1,-1\}$\;
                \marketorder$(\epsilon \cdot q)$\;
                }
            \end{algorithm}


            \begin{algorithm}
                \DontPrintSemicolon
                \caption{Value Agent trading strategy}
                \label{alg:value_agents}
                \KwData{Arrival rate $\lambda_a$, fundamental value series $\{r_t\}_{t > 0}$, spread depth $\delta_s \in \mathbb{N}$, order size $q$, inside spread probability $\xi \in [0,1]$, observation noise variance $\sigma_n$}
                \SetKwProg{valueagent}{ValueAgent}{}{end}
                \valueagent{$(\lambda_a, \{r_t\}, \delta_s, q, \xi, \sigma_n )$}{
                    \While{\marketopen$()$}{
                        $t \leftarrow $ \currenttime$()$\;
                        $\widehat{r}_t \leftarrow$ \ProgSty{update\_fundamental\_estimate}$(r_t, \sigma_n)$\;
                        $b_t, a_t \leftarrow$ \bestbid$()$, \bestask$()$\;
                        $m_t := \frac{1}{2} (a_t + b_t)$\;
                        $s_t := a_t - b_t$\;
                        Sample $X \sim \bernoulli(\xi)$\;
                        \eIf
                        {$X = 1$}
                        {
                        $\Delta_{\text{max}} \leftarrow s_t \cdot \delta_s$\;
                        Sample $\Delta \sim \udist\{0, 1, \ldots, \Delta_{\text{max}}\}$
                        }
                        {$\Delta \leftarrow 0$}
                        \eIf
                        {
                            $m_t < \widehat{r}_t$
                        }
                        {
                            $p \leftarrow b_t + \Delta$\;
                            \limitorder$(q, p)$
                        }
                        {
                            $p \leftarrow a_t - \Delta$\;
                            \limitorder$(-q, p)$
                        }
                        
                        Sample $t' \sim \expodist(\lambda_a)$\;
                        \wakeup$(t + t')$
                    }
                }
            \end{algorithm}

        \subsubsection{\textbf{Market Maker Agents:}}
        
            These play a central role in financial markets by providing liquidity allowing buyers or sellers to find a match for their orders. We implement a simple market maker in a similar fashion to that in \cite{wah2017welfare, chakraborty2011market} that utilizes both the order book information and exogenous fundamental time-series and places orders on both sides of the limit order book with varying order sizes and participation rates.

            \begin{algorithm}
                \DontPrintSemicolon
                \caption{Market Maker Agent trading strategy}
                \label{alg:market_maker_agent}
                \KwData{Fundamental value series $\{r_t\}_{t > 0}$, Max arrival rate $\lambda_a$, maximum order size $q_{\text{max}}$, maximum spread depth $\delta_s \in \mathbb{N}$}
                \SetKwProg{marketmakeragent}{MarketMakerAgent}{}{end}
                
                \marketmakeragent{$(\{r_t\}, \lambda_a, q_{\text{max}}, \delta_s)$}{
                    \While{\marketopen$()$}{
                        $t \leftarrow $ \currenttime$()$\;
                        \cancelorders$()$\;
                        $b_t, a_t \leftarrow$ \bestbid$()$, \bestask$()$\;
                        $m_t := \frac{1}{2} (a_t + b_t)$\;
                        $\widehat{r}_t \leftarrow$ \ProgSty{update\_fundamental\_estimate}$(r_t)$\;
            
                        $p_t := \frac{1}{2} (m_t + \widehat{r}_t)$ 
                        
                        sample $q_t \sim \udist\{1, \ldots, q_{\text{max}}\}$\;
                        sample $s_t \sim \udist\{1, \ldots, \delta_s\}$\;
                        
                        {
                        \limitorder$(\frac{1}{2} q_t, p_t - s_t)$\;
                        \limitorder$(-\frac{1}{2} q_t, p_t + s_t)$\; 
                        }
                        Sample $t' \sim \expodist(\lambda_a)$\;
                        \wakeup$( t + t')$\;
                    }
                }
            \end{algorithm}


        \subsubsection{\textbf{Market Taker Agents:}}
        
            These agents place large orders over a period of time consuming liquidity from the order book. In our experiments, we developed two types of market taker agents adapted from common trading practices. The first \textit{TWAP agent}, aims to execute the large order at a price as close as possible to the time weighted average price. It does this by breaking up the large order into smaller orders of equal size and dynamically placing these orders in the market using evenly divided time slots during a pre-defined time window. The second type of market taker agent, \textit{VWAP agent}, uses a similar approach. However, the size of the orders placed at each point in time is weighted against the historical traded volume of the stock. Variations of both strategies are heavily utilized in practice by trading participants wanting to buy or sell large orders on behalf of clients.

        \subsubsection{\textbf{Directional Trader Agents:}}
        
            These agents represent traders that use technical analysis to predict the direction of the stock price. In our work, we developed two simple agents; the first, \textit{mean-reversion agents}, assume that prices revert back to a mean value over a given time window. The second, \textit{momentum traders} assume that strong movement in prices in one direction are likely to be followed by movement in the same direction. Both strategies implemented compare two simple moving averages over different time periods (20 minutes and 50 minutes) and trade accordingly. In the case of the momentum agents, the agent places a buy order of a random size if the former moving average exceeds the latter and a sell order otherwise.

    \subsection{Configuration and Stylized Facts}

        Our market configuration consists of 5100 background agents (5000 noise agents and 100 value agents), 3 market makers, 6 market takers (3 TWAP agents and 3 VWAP agents) and 10 directional traders (5 momentum traders and 5 mean-reversion traders). We utilize the simulator's variable network communication properties and use a default computational delay of 50 nanoseconds and a pairwise latency between the agents and the exchange equivalent to the distance between New York and Seattle. In our experiments, we use historical mid-prices of J.P. Morgan Chase's as the fundamental value the agents observe and run the simulation across one week worth of NASDAQ data between 09:30 AM and 16:00 PM which are the market open and close times of this exchange.
        
\begin{figure}[H]
             \centering
             \subfloat[One-minute log returns]{\includegraphics[width=0.23\textwidth]{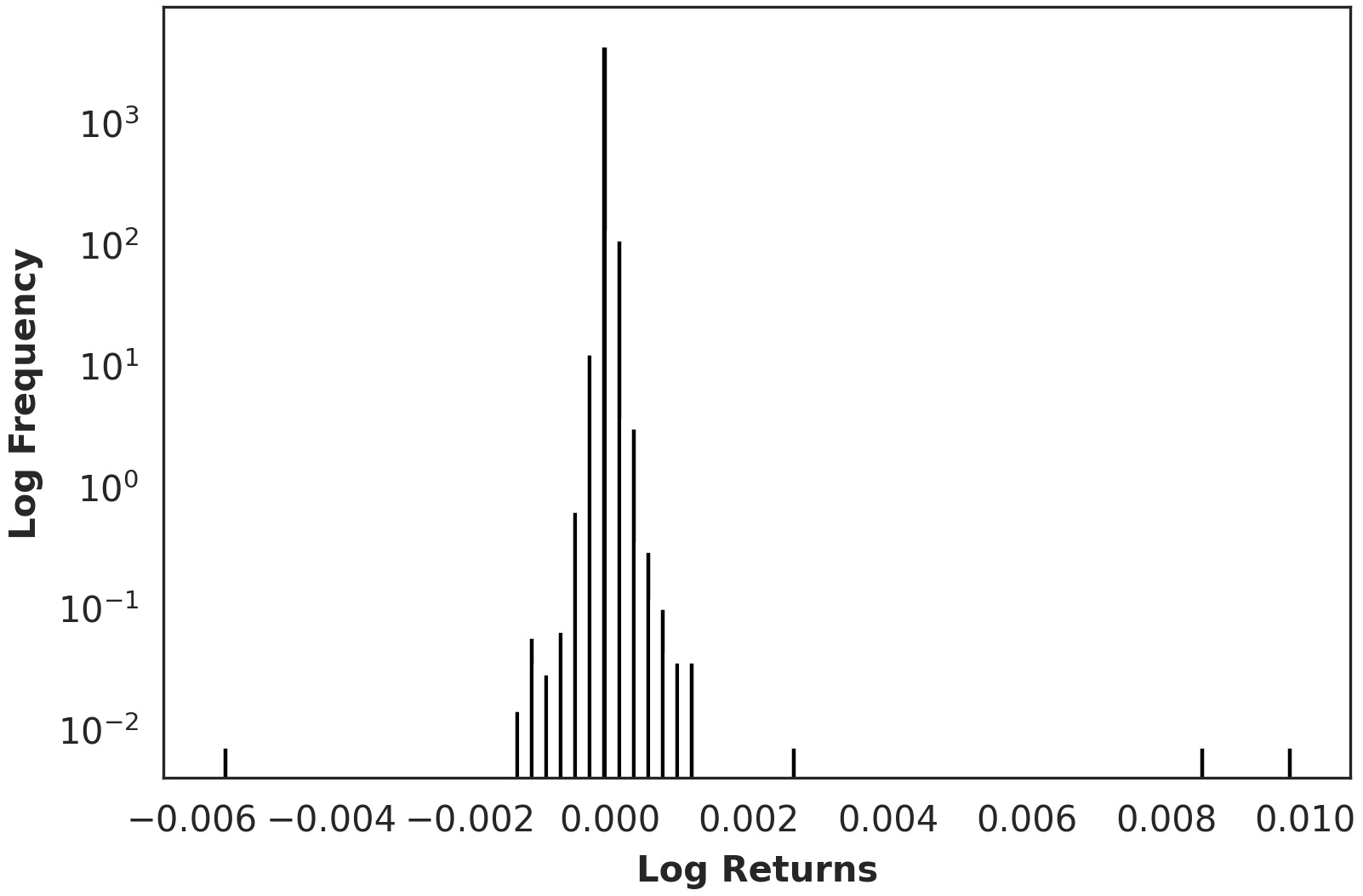}}\label{fig:one_min_returns}
             \hfill
             \subfloat[Ten-minute log returns]{\includegraphics[width=0.23\textwidth]{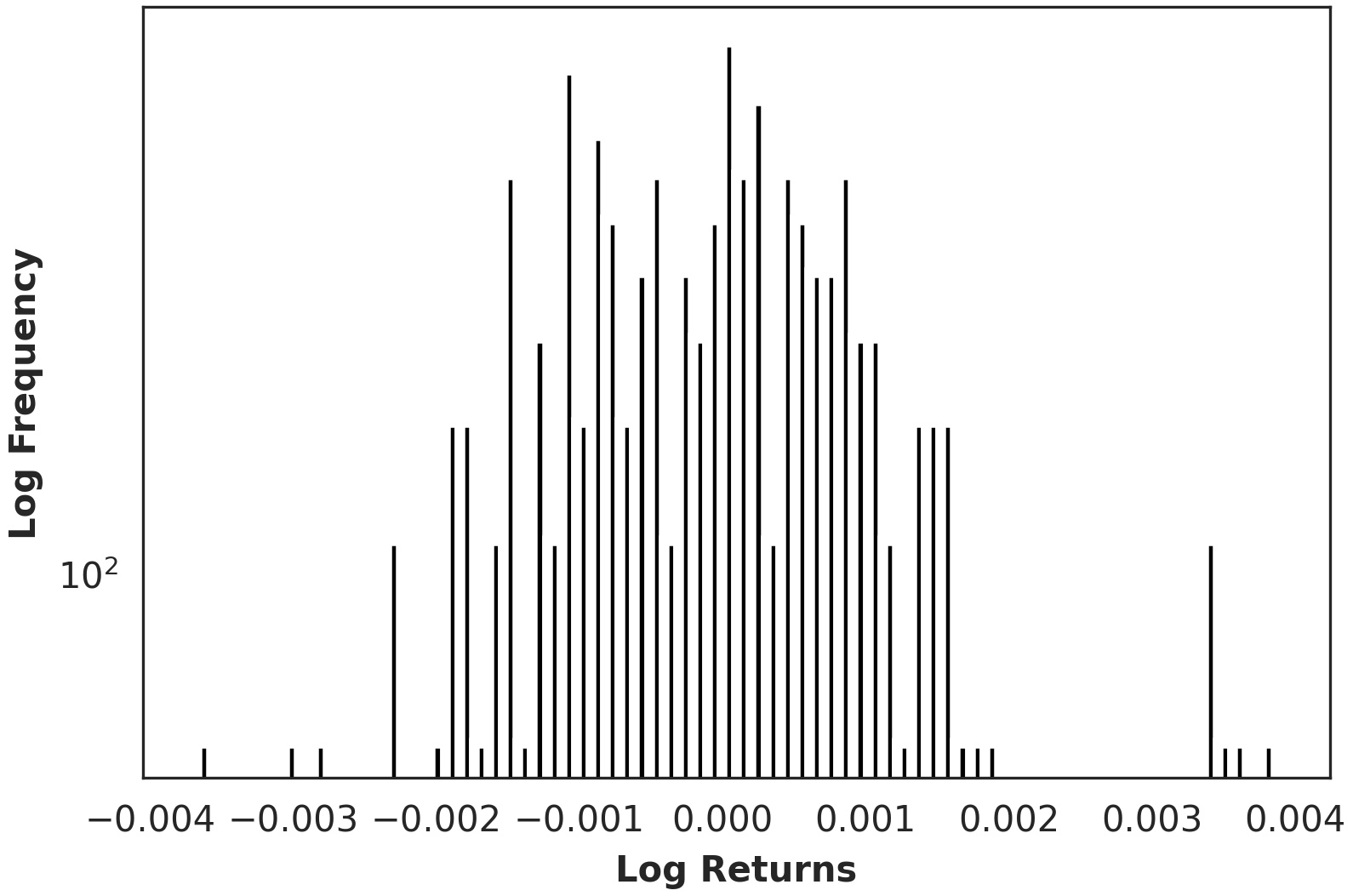}}\label{fig:ten_min_returns}
             \caption{Log returns distribution}
        \end{figure}

        In order to assess the realism of our simulation, we compare a number of its' statistical properties (referred to as \emph{stylized facts}) to those observed in real markets. \cite{bouchaud2018trades} provides a detailed overview of these statistical properties covering stylized facts about asset price return distributions, order volumes and order flow. Figures \ref{fig:one_min_returns} and \ref{fig:ten_min_returns} show the distribution of daily asset price returns for the simulation dataset we used with the minutely returns displaying fat tails that become slimmer when the period over which the returns are calculated is increased. This matches the statistical properties observed in real market data according to \cite{bouchaud2018trades}.
        
%% file: body/_5_classifying_trading_agents.tex
\section{Classifying Trading Agent Archetypes} \label{sec:om}

    \subsection{Overview}

        We formulate our approach to opponent modeling in a limit order book as a supervised learning problem of identifying (classifying) agent \emph{archetypes} explicitly programmed in the simulator, a task which can be cast in the auxiliary tasks framework \cite{jaderberg2016reinforcement, hernandez2019agent} for semi-supervised reinforcement learning. Using a classification approach for opponent modeling \cite{weber2009data} allows for predicting properties of the other agents as opposed to predicting the future actions of the agents, which can be useful in a range of market scenarios discussed in our work. 

    \subsection{Model}
        
\begin{figure}[!htb]
             \centering
             \includegraphics[width=0.13\textwidth]{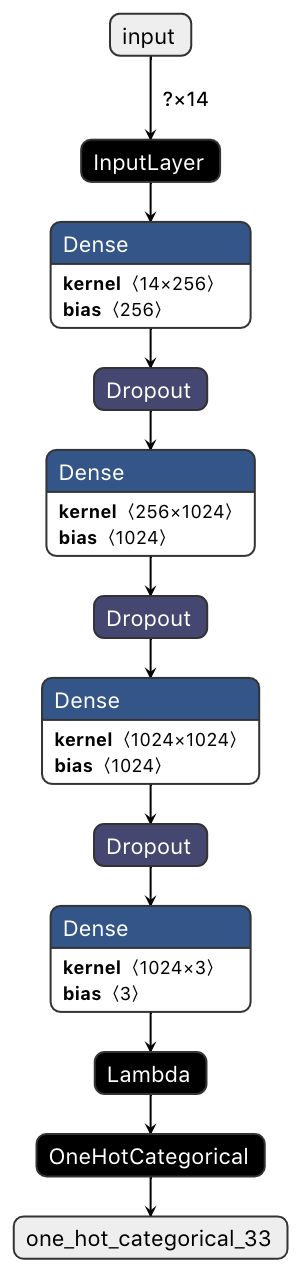}
             \caption{Neural network model architecture diagram}
             \label{fig:nn_arch}
        \end{figure}

        The classification model used is a simple feed-forward neural network (see figure \ref{fig:nn_arch}) that attempts to infer the agent archetype given the limit order book state and the agent’s order. The state of the limit order book is given by the z-score normalised ask and bid prices and volumes for the first 5 price levels. The agent's order is represented by the direction of the order (buy or sell), its price and size. The model consists of 4 hidden layers with 256, 1024, 1024 and 1024 units respectively. In order to reduce over-fitting on the training data, we add a drop out layer after each hidden layer. We used a mixture of ReLU and sigmoid activation functions, an ADAM optimizer to update the network weights based on the training data and ran the model for 200 epochs. Finally, we used an open-source hyper-parameter optimization framework \cite{bergstra2013making} to find the best parameters to use for our model and dataset.

        \begin{table*}[!tb]
        \small
            \begin{tabular}{c | c c c | c c c | c c c | c c c} 
                \toprule
                 & 
                \multicolumn{3}{c}{\textbf{Market Maker}} & \multicolumn{3}{c}{\textbf{Market Taker}} & \multicolumn{3}{c}{\textbf{Directional Trader}} & \multicolumn{3}{c}{\textbf{Background Agent}} \\
                 & Precision & Recall & F-score & Precision & Recall & F-score  & Precision & Recall & F-score  & Precision & Recall & F-score \\
                 \hline
                 KNN & 0.55 & 0.69 & 0.61 & 0.47 & 0.66 & 0.55 & 0.41 & 0.29 & 0.34 & 0.41 & 0.26 & 0.32 \\
                 \hline
                 Linear SVM & 0.73 & 0.56 & 0.64 & 0.36 & 0.24 & 0.29 & 0.41 & 0.25 & 0.31 & 0.31 & 0.61 & 0.41 \\
                 \hline
                 RBF SVM & 0.51 & 0.80 & 0.63 & 0.61 & 0.54 & 0.57 & 0.42 & 0.35 & 0.38 & 0.44 & 0.34 & 0.38 \\
                 \hline
                 Decision Tree & 0.99 & 0.47 & 0.64 & 0.69 & 0.70 & 0.70 & 0.55 & 0.85 & 0.67 & 0.67 & 0.64 & 0.65 \\
                 \hline
                 Random Forest & 0.44 & 0.61 & 0.51 & 0.40 & 0.65 & 0.50 & 0.38 & 0.25 & 0.30 & 0.53 & 0.21 & 0.30 \\
                 \hline
                 AdaBoost & 0.80 & 0.59 & 0.68 & 0.78 & 0.69 & 0.73 & 0.57 & 0.84 & 0.68 & 0.70 & 0.63 & 0.67 \\
                 \hline
                 Naive Bayes & 0.94 & 0.47 & 0.63 & 0.18 & 0.03 & 0.06 & 0.41 & 0.92 & 0.56 & 0.57 & 0.57 & 0.57 \\ 
                 \hline
                 \textbf{Our model} & 0.76 & 0.75 & 0.75 & 0.70 & 0.71 & 0.71 & 0.60 & 0.63 & 0.61 & 0.68 & 0.63 & 0.65 \\
                 \bottomrule
            \end{tabular}
            \label{tab:tab_om_results}
            \caption{Opponent modelling classification results}
        \end{table*}
        
        \begin{figure*}[!tb]
            \centering
            \begin{subfigure}[b]{0.3305\textwidth}
                \centering
                \includegraphics[width=\textwidth]{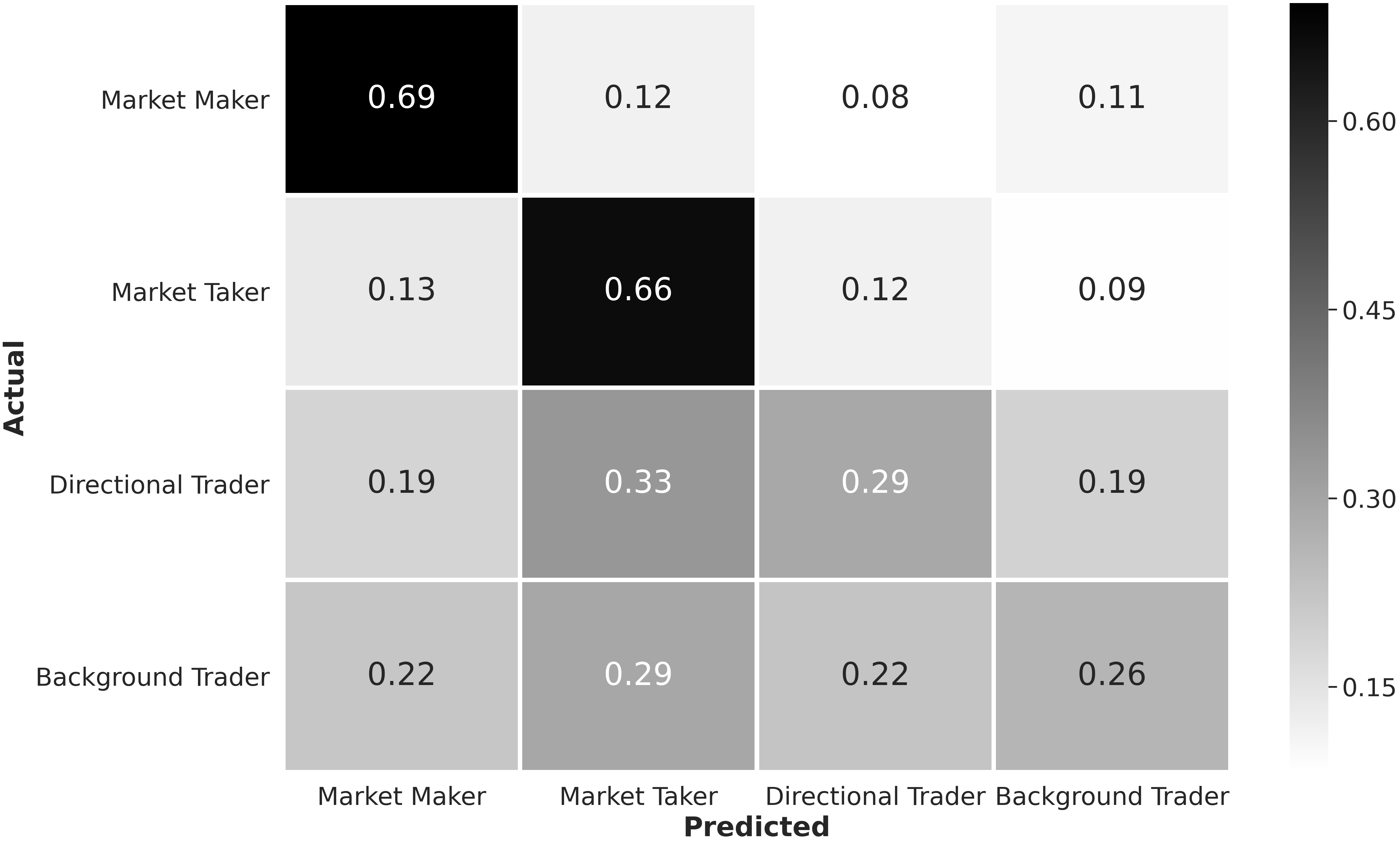}
                \caption{KNN}
            \end{subfigure}
            \begin{subfigure}[b]{0.3305\textwidth}
                \centering
                \includegraphics[width=\textwidth]{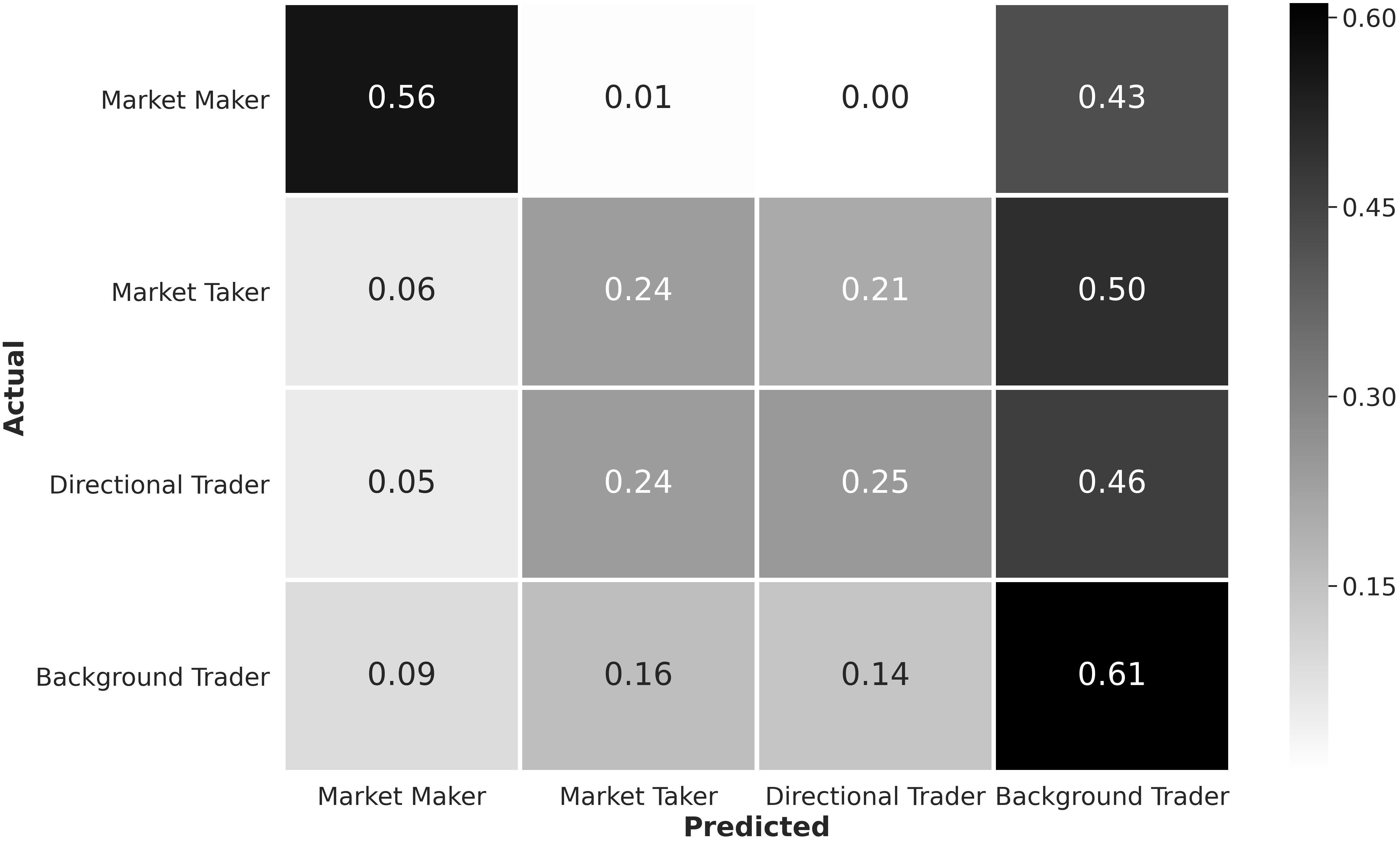}
                \caption{Linear SVM}
            \end{subfigure}
            \begin{subfigure}[b]{0.3305\textwidth}
                \centering
                \includegraphics[width=\textwidth]{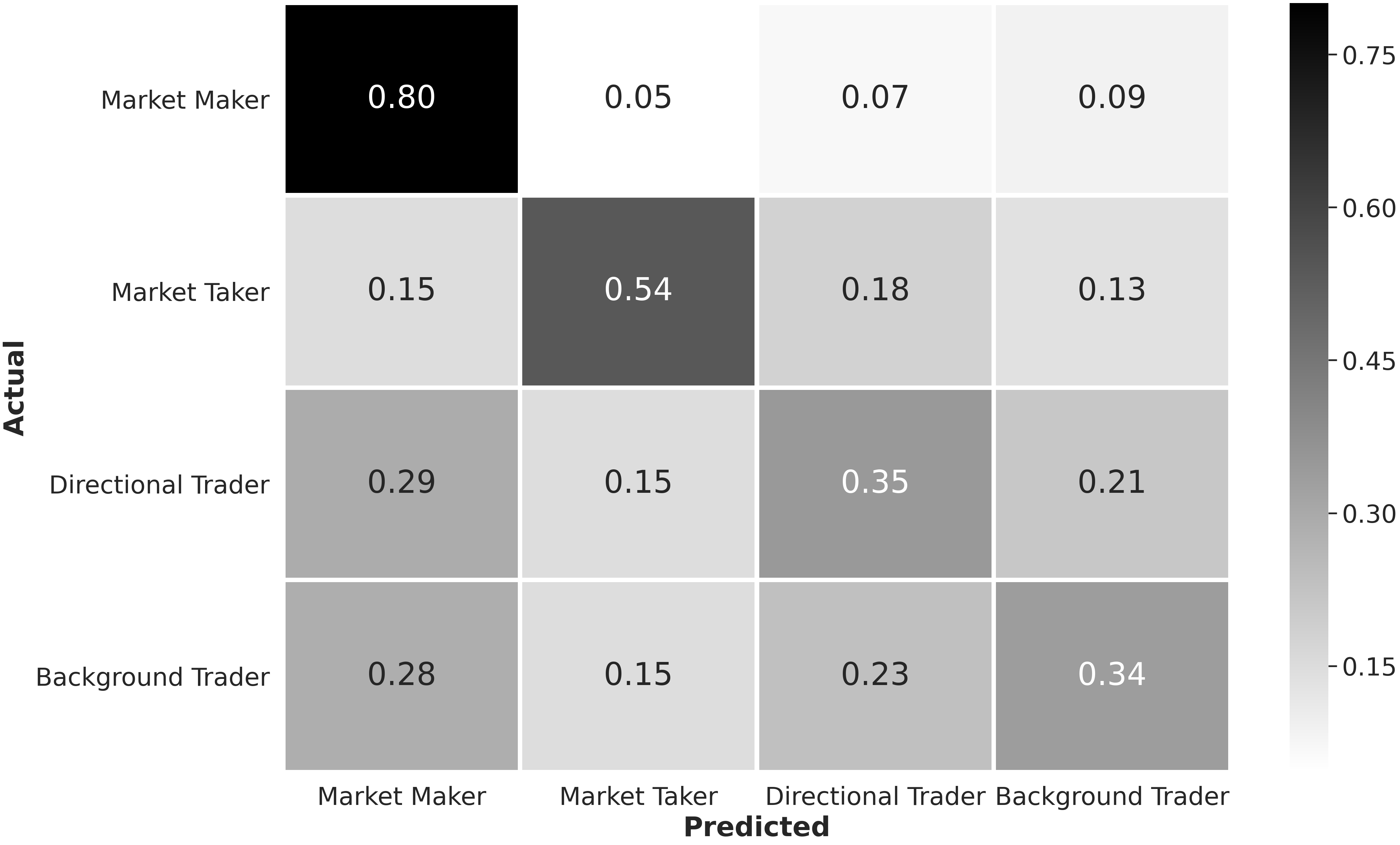}
                \caption{RBF SVM}
            \end{subfigure}
            \begin{subfigure}[b]{0.3305\textwidth}
                \centering
                \includegraphics[width=\textwidth]{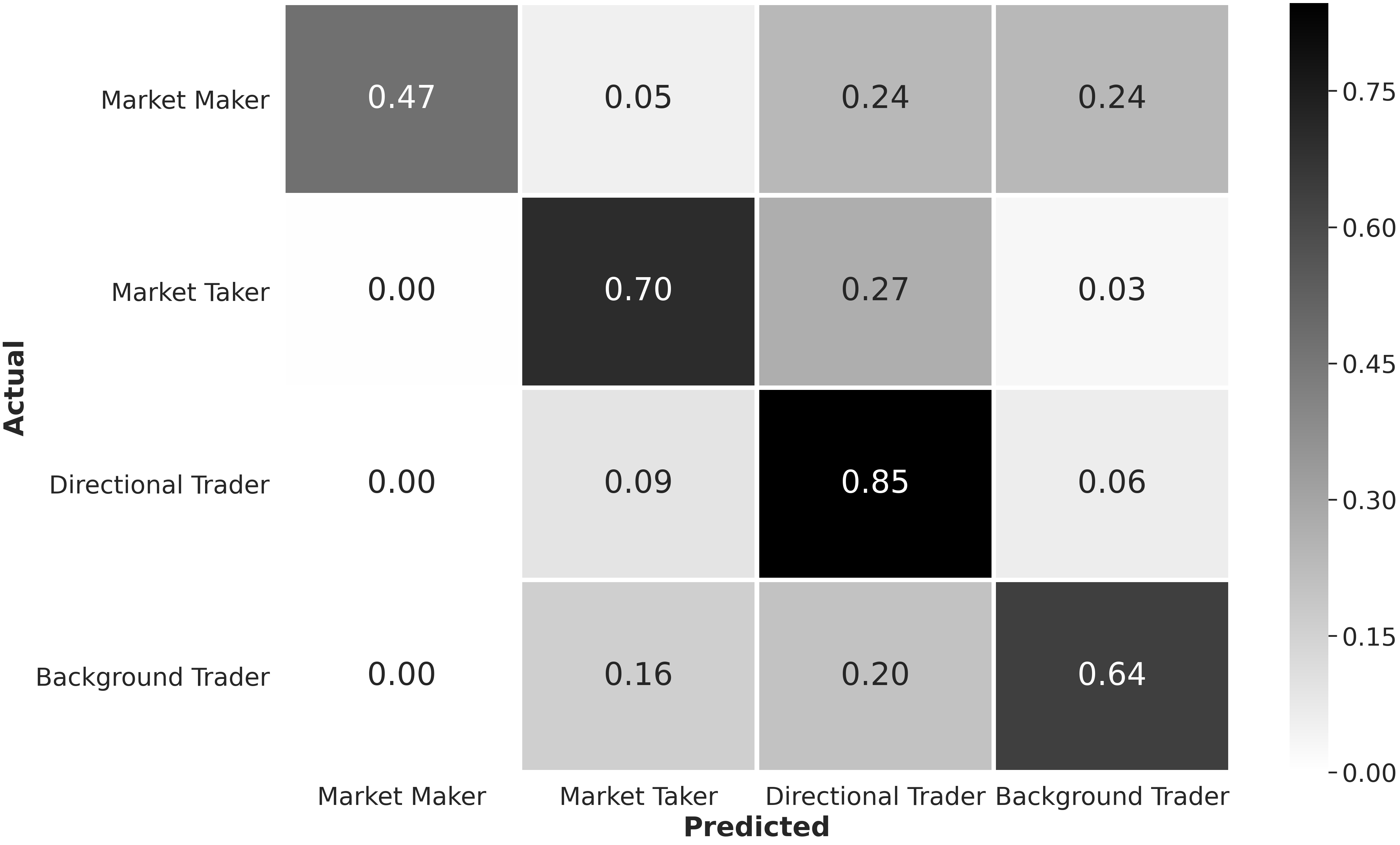}
                \caption{Decision Tree}
            \end{subfigure}
            \begin{subfigure}[b]{0.3305\textwidth}
                \centering
                \includegraphics[width=\textwidth]{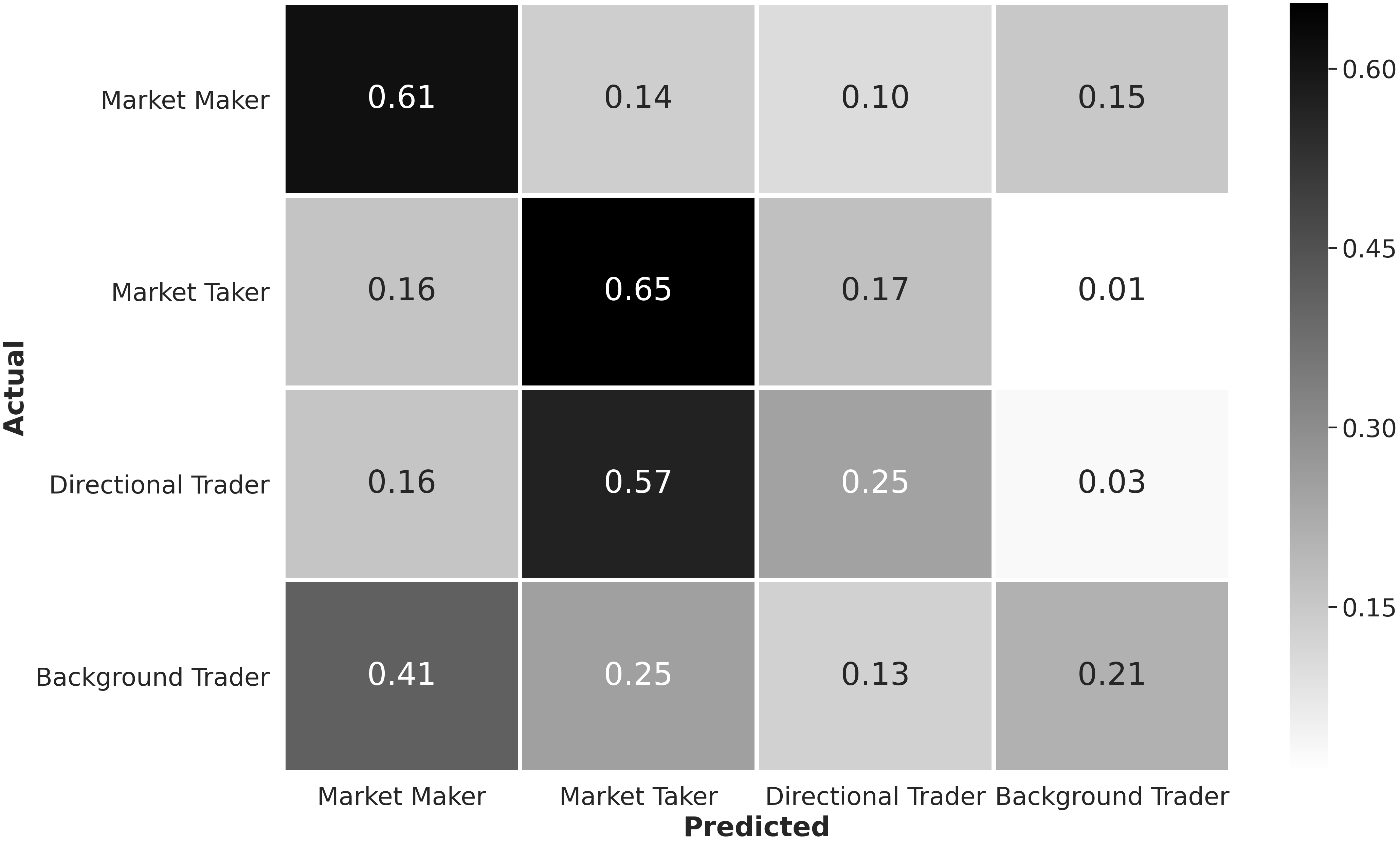}
                \caption{Random Forest}
            \end{subfigure}
            \begin{subfigure}[b]{0.3305\textwidth}
                \centering
                \includegraphics[width=\textwidth]{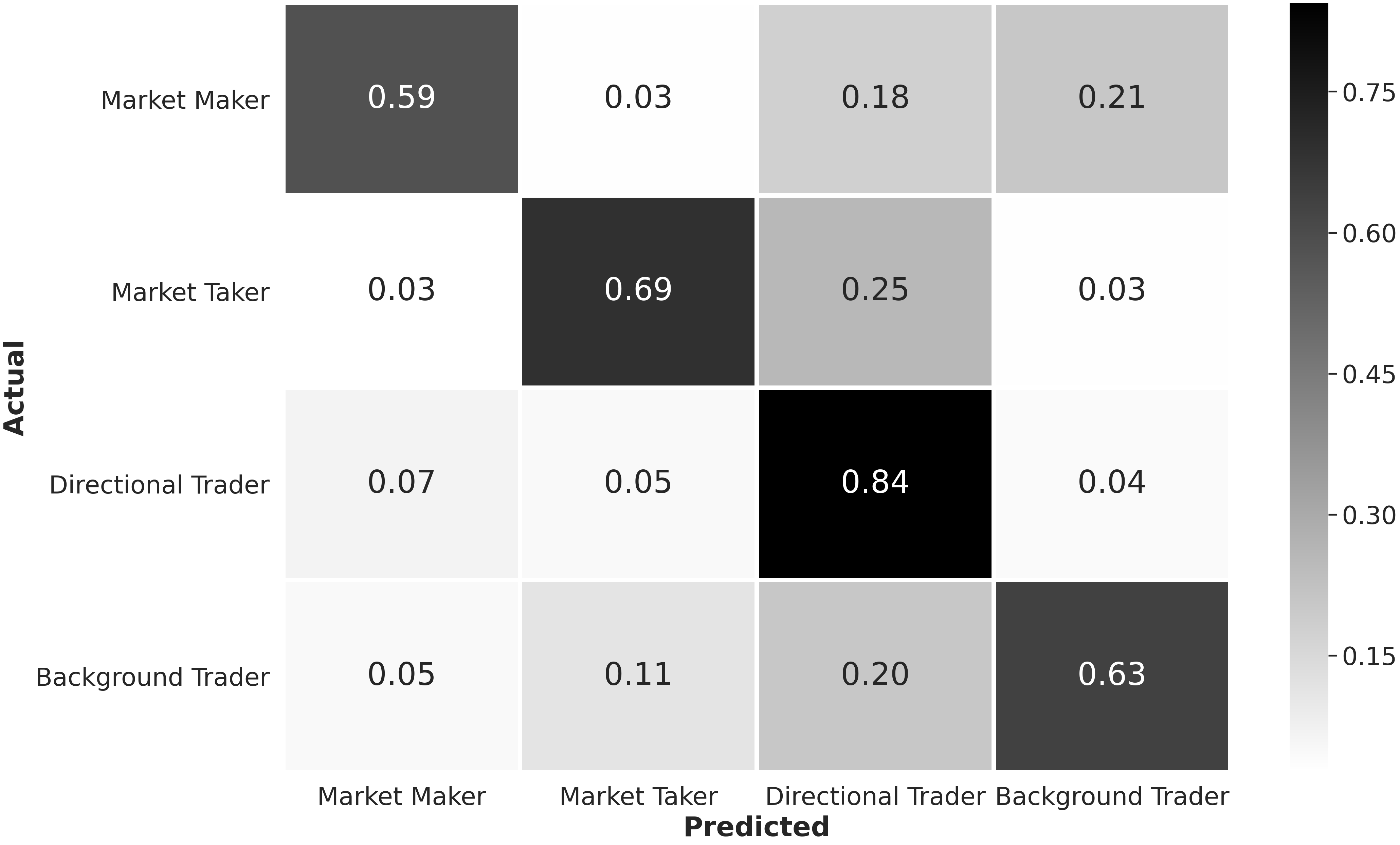}
                \caption{AdaBoost}
            \end{subfigure}
            \begin{subfigure}[b]{0.3305\textwidth}
                \centering
                \includegraphics[width=\textwidth]{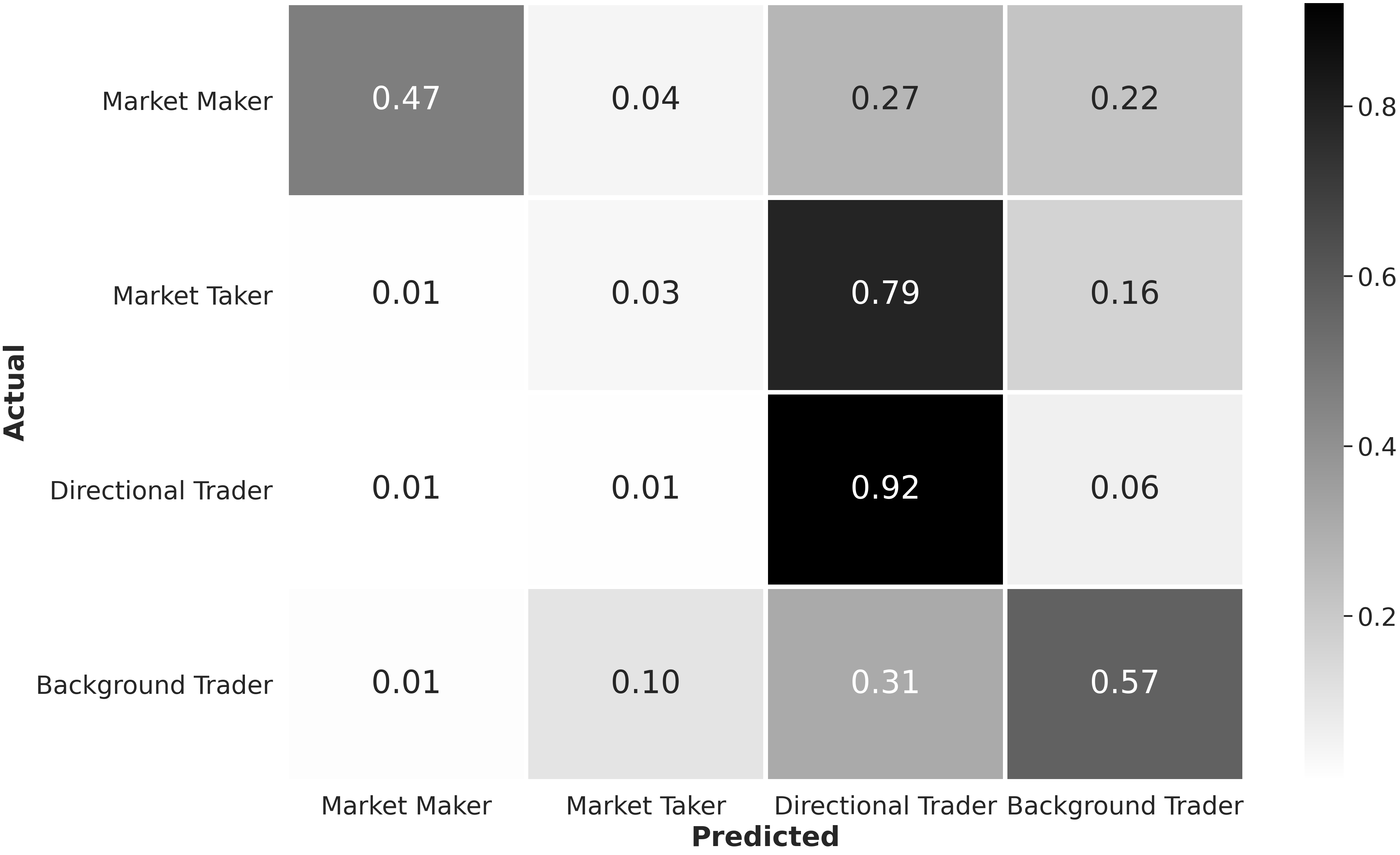}
                \caption{Naive Bayes}
            \end{subfigure}
            \begin{subfigure}[b]{0.3305\textwidth}
                \centering
                \includegraphics[width=\textwidth]{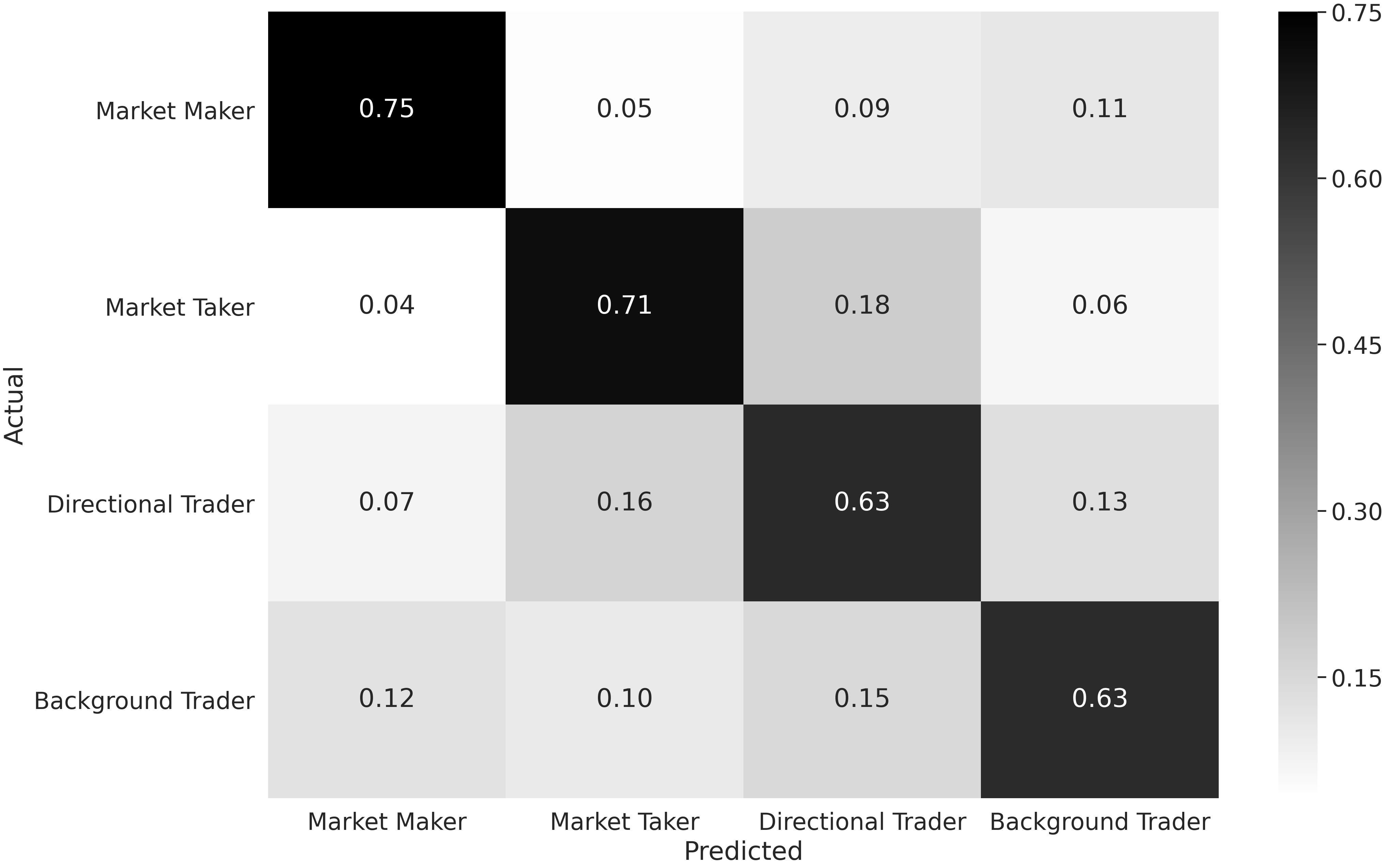}
                \caption{Our model}
            \end{subfigure}
            \label{fig:fig_om_results}
            \caption{Baseline classification models confusion matrices}
        \end{figure*}

    \subsection{Experiments and Results}
    
        We run our ABM using fundamental values covering five trading days for one stock and produce labelled synthetic market data that we use offline for the opponent classification exercise. An issue arises regarding the different time scales that each agent archetype operates on. For example, some archetypes place substantially more orders than others, leading to the data set we generate having high imbalance between the classes. We produce a balanced dataset by simply down-sampling the over-represented classes.
        
        We train our classifier on data for 3 days and evaluate its performance on the subsequent 2 days. Figure 6 shows the confusion matrix for the four trading archetypes in our ABM model. We compare the performance of our model against a range of classifiers and show the precision, recall and F1-score associated with each agent archetype in table 1.
        
        The confusion matrix for our model shows relatively good performance when looking at the true positives. Comparing the precision, recall and F1-scores of our model against the baseline models shown in table 1, we see mixed performance indicating an ensemble model is most suited. Our model performs best in identifying the market makers and market takers and displays lower performance identifying the directional and background traders. It is important to note that the down-sampling of the data and the lack of long term memory in the neural network architecture plays a significant part in this. For example, the directional traders developed rely on past observations of the order book to place their orders that our model and data sampling methodology makes worse. On the other hand, we see the ensemble AdaBoost model \cite{hastie2009multi} performing well confirming our hypothesis about the need for an ensemble model. The AdaBoost model \cite{hastie2009multi}, however, does not show the same performance for the market makers compared to our model.

%% file: body/_6_behavioural_cloning.tex
\section{Behavioural Cloning of Trading Agent Archetypes} \label{sec:bc}
    \subsection{Overview}
    
        Using the same labelled synthetic data produced by the ABM model, we formulate our approach to behavioural cloning of the agents as a supervised regression problem of learning the agents' order price and size given observations of the limit order book. This approach ties well with the paradigm taken in the previous classification exercise and allows us to use the same dataset but change the inputs and labels of our model.

    \subsection{Model}
        We use the same neural network architecture shown in figure \ref{fig:nn_arch} but change the last layer of the neural network to output a vector of z-score normalized order prices and sizes. Instead of training the model on the entire dataset, we train it only on the observations and actions of each agent archetype. In this model, we also used the hyper-parameter optimization framework \cite{bergstra2013making} to find the best parameters to use for our model and dataset for each agent archetype.

    \subsection{Experiments and Results}
\begin{figure*}[!tb]
            \centering
            \begin{subfigure}[b]{0.48\textwidth}
                \centering
                \includegraphics[width=\textwidth]{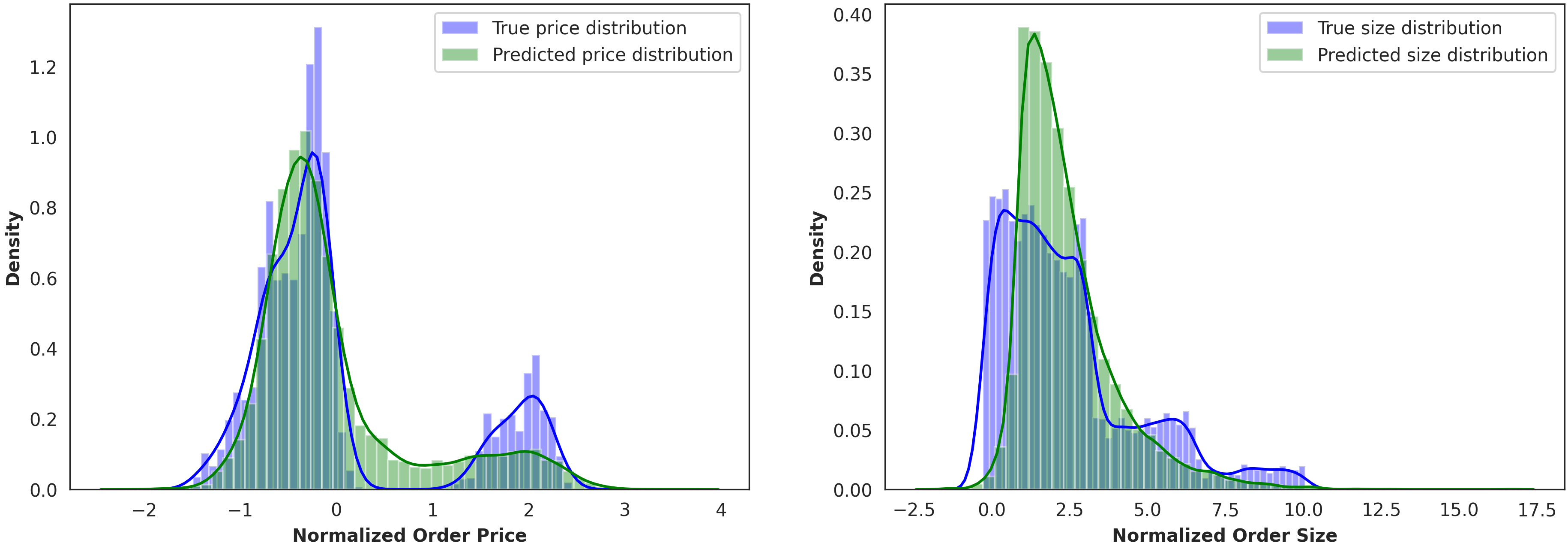}
                \caption{Market Makers}
                \label{fig:bc_mm}
            \end{subfigure}
            \begin{subfigure}[b]{0.48\textwidth}
                \centering
                \includegraphics[width=\textwidth]{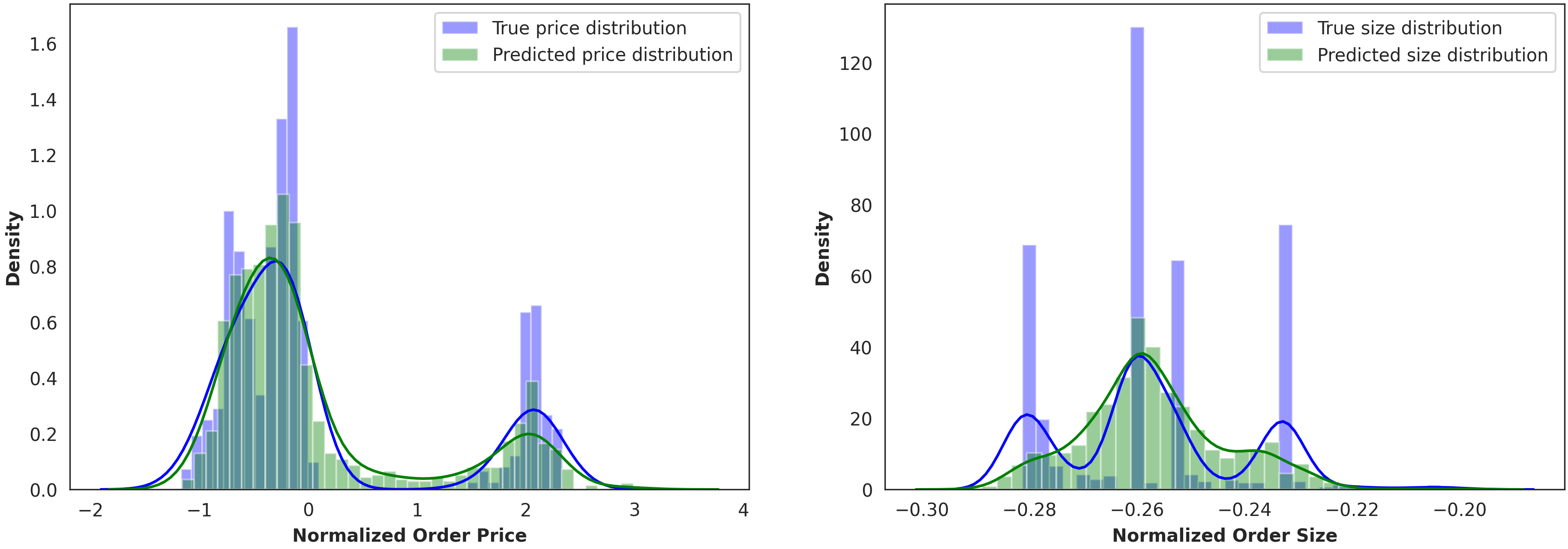}
                \caption{Market Takers}
                \label{fig:bc_mt}
            \end{subfigure}
            \begin{subfigure}[b]{0.48\textwidth}
                \centering
                \includegraphics[width=\textwidth]{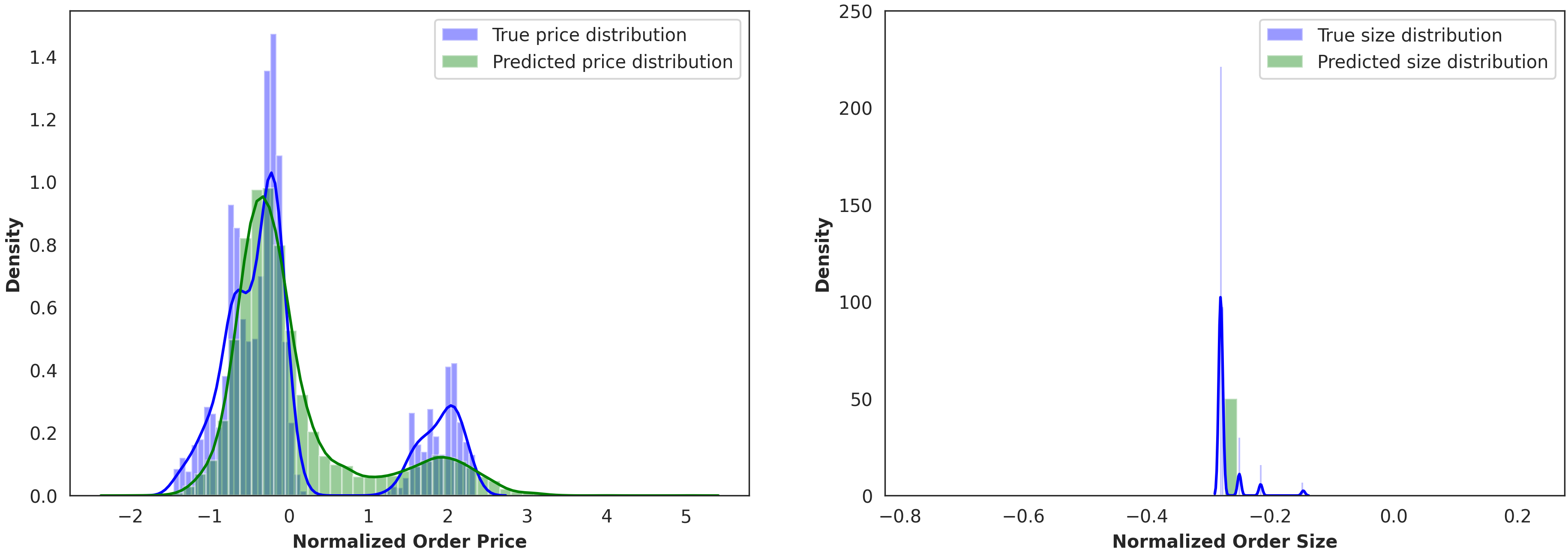}
                \caption{Directional Traders}
                \label{fig:bc_dt}
            \end{subfigure}
            \begin{subfigure}[b]{0.48\textwidth}
                \centering
                \includegraphics[width=\textwidth]{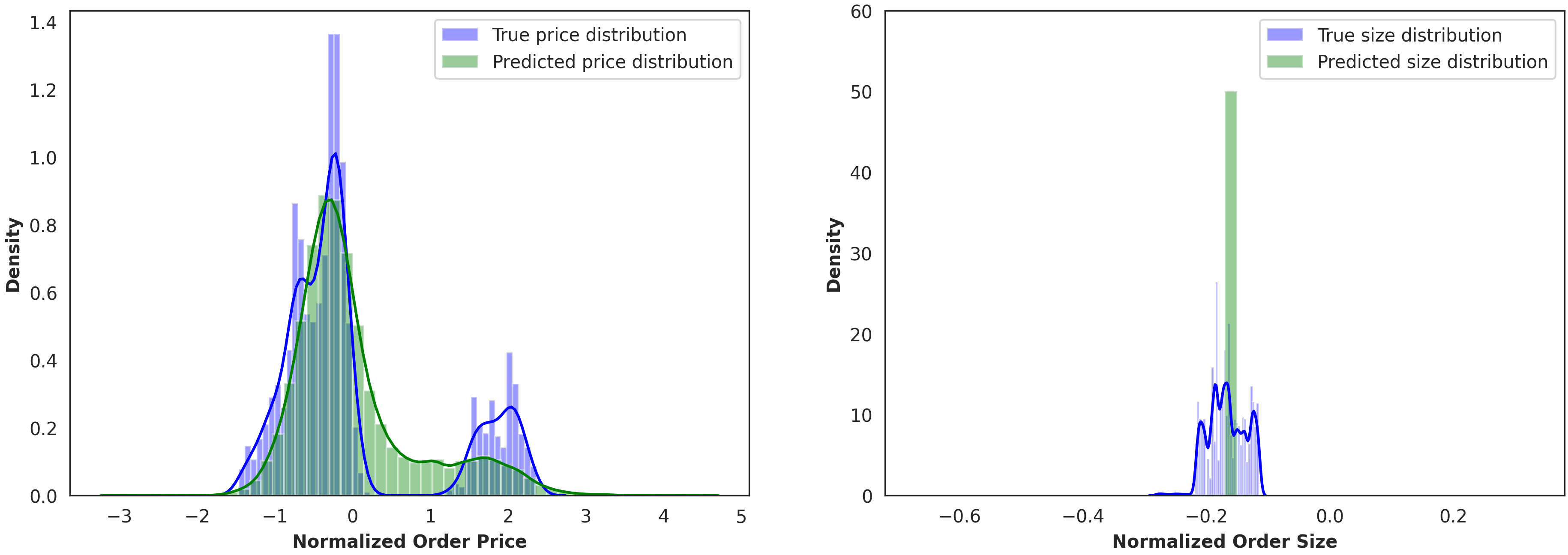}
                \caption{Background Agents}
                \label{fig:bc_bt}
            \end{subfigure}
            \caption{Distribution of order prices and sizes for the cloned agent archetypes}
        \end{figure*}

        We train our regression model on data specific to each agent archetype for 3 days and evaluate its performance on the subsequent 2 days. Our results indicate the possibility of predicting the order prices and sizes with room for improvement. Figures \ref{fig:bc_mm}, \ref{fig:bc_mt}, \ref{fig:bc_dt}, \ref{fig:bc_bt} show the univariate distributions of the predicted and ground truth data of both order prices and sizes as well as a kernel density estimate for better visualisation. In almost all cases, we see non symmetric distributions suggesting the need for a more complex regression model.
        
%% file: body/_7_conclusion_future_work.tex
\section{Conclusion} \label{sec:conclusion_future_work}

    We have explored and shown the feasibility of applying opponent modelling and behavioural cloning techniques in simulated continuous double auctions. To this end, we built an agent-based model consisting of four different trading agent archetypes following strategies common in practice. We assessed the realism of our ABM against known statistical properties of real limit order book data to ensure our simulation is realistic. We formulated our approach to opponent modeling in a limit order book as a supervised learning problem of identifying (classifying) agent archetypes explicitly designed in the ABM and designed a neural network model that performs relatively well against a number of baseline methods. Finally, we used the same data and showed the feasibility of performing behavioural cloning by using a similar regression model to perform this task.
    
    The simple experiments explored and the initial results obtained are encouraging evidence that opponent modeling and behavioural cloning methods are potentially feasible and useful in settings of interest to the different actors in the financial markets. Future work in this direction could explore more unsupervised or semi-supervised learning approaches for opponent modelling. However, the challenges anticipated following such approaches would be around labeling any clusters obtained which would still require domain knowledge. This suggests the need for both supervised and unsupervised learning approaches.
    
    Methods of opponent modeling in financial multi-agent systems can be useful in a number of scenarios and from different perspectives, including but not limited to:
    
    \textbf{Regulation and Safety:}
    Market manipulation strategies represent an element of price distortion and creation of artificial market conditions, which in turn impedes market efficiency, transparency and fairness. In the limit order book context, spoofing agents, for instance, submit orders only to cancel them before their execution and then submit a buy or sell order to benefit from the change that their initial (fake) order caused. The ability to identify these spoofing agents and explicitly model their policies could help in the early detection of market abuse and avoiding adverse consequences like the flash crash that happened on May 6, 2010, causing a large and quickly reverting dip in the Dow Jones Industrial Average stocks. Moreover, \textit{Level 4} information, consisting of LOB labeled records, are exclusively available to exchanges and regulators to verify that market participants are complying with rules and processes \cite{paddrik2017effects}. Having access to such a data set is an opportunity to make a more informed and dynamic utility modeling. Quantifying proportions of market makers, liquidity takers and buy side participants would be valuable in detecting signs of recession or unhealthy market conditions.
    
    \textbf{Decision Making and Automation:}
    Market participants usually only have access to (at most) \textit{Level 3} LOB data \cite{paddrik2017effects}, which is \textit{anonymized} order flow data with new, modified, and canceled limit orders placed at a given snapshot. Although this poses a challenge to the disambiguation of limit orders placed in the order book because it imposes a myopic approach as opposed to using a consistent timeseries of consecutive orders placed by the same agent, unlabeled data does not fall short of information and opportunities. Opponent modeling can be used for the detection of iceberg orders and other forms of hidden liquidity \cite{hautsch2012dark}. By studying the occurrence of such orders, a \textit{posteriori}, in the LOB and the surrounding market conditions, we can attempt to model the utility behind such strategies. \cite{hautsch2012dark} studied the correlation between dark orders and market conditions and showed that market conditions reflected by the bid-ask spread, depth and price movements significantly impact the aggressiveness of `dark' liquidity supply. If a trader can detect such dark orders through recognizing a pattern that iceberg order placing agents fail to conceal or through modeling the utility of such agents and the specific cases that make them submit such orders, they can estimate and take into account these orders' effect on the LOB balance and predict a larger than observed spread.  
    
    Additionally, the Request for Quotes (RFQ) process is another setting where we could benefit from opponent modeling. In this setting, clients send quote requests to multiple dealers for a specific quantity of a security. Ultimately, the dealer with the most attractive offer wins the auction. In this single, reverse, and static auction setting, one buyer (the auctioneer/the client) is looking to buy a single security from multiple sellers (the dealers) who do not observe the offers submitted by the other dealers, and only receive feedback from the auctioneer once the latter's decision for the winner of the auction is made. It is therefore important for dealers receiving RFQs from different clients to have a good estimate of the expected level at which the other dealers are willing to sell the security to avoid mis-evaluating the offer price. This can in theory be accomplished by having a good estimate of the fair value of the asset in question. However, it is difficult to estimate the fair value of the asset if the asset is illiquid. If a participant is able to infer the utilities, and by extension, the expected offer prices of each of the other dealers in the auction, they can narrow down their uncertainty about the fair value of the asset and avoid mispricing it. In RFQs, apart from modeling competitors, it can be extremely valuable to be able to model clients. In this case data is labeled, as sell side firms (banks) are aware of their RFQ counter party (client). Performing efficient client tiering would allow to respond to client orders in an efficient manner, which is beneficial for both the client and the bank.

    Behavioural cloning, on the other hand, is very useful in trading applications where agents are able to demonstrate their behaviour rather than specify their policy directly or formulate their reward. The abundance of trading agent demonstration data stored by financial institutions makes this paradigm very interesting to explore and presents a unique opportunity to use machine learning for both regular and algorithmic trading applications.